\documentclass[twocolumn]{aastex62}
\submitjournal{The Astrophysical Journal Letters}
\usepackage{graphicx}
\usepackage{float}
\usepackage{amsmath}
\usepackage{lineno}
\usepackage[caption=false]{subfig}
\usepackage[utf8]{inputenc}

\shorttitle{TeV emission and powerful pulsars}
\shortauthors{HAWC Collaboration}

\begin{document}
\title{Evidence that Ultra-High-Energy Gamma Rays are a Universal Feature Near Powerful Pulsars}

\author{A.~Albert}
\affiliation{Physics Division, Los Alamos National Laboratory, Los Alamos, NM, USA }
\author{R.~Alfaro}
\affiliation{Instituto de F\'{i}sica, Universidad Nacional Autónoma de México, Ciudad de Mexico, Mexico }
\author{C.~Alvarez}
\affiliation{Universidad Autónoma de Chiapas, Tuxtla Gutiérrez, Chiapas, México}
\author{J.D.~Álvarez}
\affiliation{Universidad Michoacana de San Nicolás de Hidalgo, Morelia, Mexico }
\author{J.R.~Angeles Camacho}
\affiliation{Instituto de F\'{i}sica, Universidad Nacional Autónoma de México, Ciudad de Mexico, Mexico }
\author{J.C.~Arteaga-Velázquez}
\affiliation{Universidad Michoacana de San Nicolás de Hidalgo, Morelia, Mexico }
\author{K.P.~Arunbabu}
\affiliation{Instituto de Geof\'{i}sica, Universidad Nacional Autónoma de México, Ciudad de Mexico, Mexico }
\author{D.~Avila Rojas}
\affiliation{Instituto de F\'{i}sica, Universidad Nacional Autónoma de México, Ciudad de Mexico, Mexico }
\author{H.A.~Ayala Solares}
\affiliation{Department of Physics, Pennsylvania State University, University Park, PA, USA }
\author{V.~Baghmanyan}
\affiliation{Institute of Nuclear Physics Polish Academy of Sciences, PL-31342 IFJ-PAN, Krakow, Poland }
\author{E.~Belmont-Moreno}
\affiliation{Instituto de F\'{i}sica, Universidad Nacional Autónoma de México, Ciudad de Mexico, Mexico }
\author{S.Y.~BenZvi}
\affiliation{Department of Physics \& Astronomy, University of Rochester, Rochester, NY , USA }
\author{C.~Brisbois}
\affiliation{Department of Physics, University of Maryland, College Park, MD, USA }
\author{K.S.~Caballero-Mora}
\affiliation{Universidad Autónoma de Chiapas, Tuxtla Gutiérrez, Chiapas, México}
\author{T.~Capistrán}
\affiliation{Instituto de Astronom\'{i}a, Universidad Nacional Autónoma de México, Ciudad de Mexico, Mexico }
\author{A.~Carramiñana}
\affiliation{Instituto Nacional de Astrof\'{i}sica, Óptica y Electrónica, Puebla, Mexico }
\author{S.~Casanova}
\affiliation{Institute of Nuclear Physics Polish Academy of Sciences, PL-31342 IFJ-PAN, Krakow, Poland }
\author{U.~Cotti}
\affiliation{Universidad Michoacana de San Nicolás de Hidalgo, Morelia, Mexico }
\author{J.~Cotzomi}
\affiliation{Facultad de Ciencias F\'{i}sico Matemáticas, Benemérita Universidad Autónoma de Puebla, Puebla, Mexico }
\author{S.~Coutiño de León}
\affiliation{Instituto Nacional de Astrof\'{i}sica, Óptica y Electrónica, Puebla, Mexico }
\author{E.~De la Fuente}
\affiliation{Departamento de F\'{i}sica, Centro Universitario de Ciencias Exactase Ingenierias, Universidad de Guadalajara, Guadalajara, Mexico }
\author{C.~de León}
\affiliation{Universidad Michoacana de San Nicolás de Hidalgo, Morelia, Mexico }
\author{R.~Diaz Hernandez}
\affiliation{Instituto Nacional de Astrof\'{i}sica, Óptica y Electrónica, Puebla, Mexico }
\author{B.L.~Dingus}
\affiliation{Physics Division, Los Alamos National Laboratory, Los Alamos, NM, USA }
\author{M.A.~DuVernois}
\affiliation{Department of Physics, University of Wisconsin-Madison, Madison, WI, USA }
\author{M.~Durocher}
\affiliation{Physics Division, Los Alamos National Laboratory, Los Alamos, NM, USA }
\author{J.C.~Díaz-Vélez}
\affiliation{Departamento de F\'{i}sica, Centro Universitario de Ciencias Exactase Ingenierias, Universidad de Guadalajara, Guadalajara, Mexico }
\author{R.W.~Ellsworth}
\affiliation{Department of Physics, University of Maryland, College Park, MD, USA }
\author{K.~Engel}
\affiliation{Department of Physics, University of Maryland, College Park, MD, USA }
\author{C.~Espinoza}
\affiliation{Instituto de F\'{i}sica, Universidad Nacional Autónoma de México, Ciudad de Mexico, Mexico }
\author{K.L.~Fan}
\affiliation{Department of Physics, University of Maryland, College Park, MD, USA }
\author{M.~Fernández Alonso}
\affiliation{Department of Physics, Pennsylvania State University, University Park, PA, USA }
\author{N.~Fraija}
\affiliation{Instituto de Astronom\'{i}a, Universidad Nacional Autónoma de México, Ciudad de Mexico, Mexico }
\author{A.~Galván-Gámez}
\affiliation{Instituto de Astronom\'{i}a, Universidad Nacional Autónoma de México, Ciudad de Mexico, Mexico }
\author{J.A.~García-González}
\affiliation{Tecnologico de Monterrey, Escuela de Ingenier\'{i}a y Ciencias, Ave. Eugenio Garza Sada 2501, Monterrey, N.L., Mexico, 64849}
\author{F.~Garfias}
\affiliation{Instituto de Astronom\'{i}a, Universidad Nacional Autónoma de México, Ciudad de Mexico, Mexico }
\author{G.~Giacinti}
\affiliation{Max-Planck Institute for Nuclear Physics, 69117 Heidelberg, Germany}
\author{M.M.~González}
\affiliation{Instituto de Astronom\'{i}a, Universidad Nacional Autónoma de México, Ciudad de Mexico, Mexico }
\author{J.A.~Goodman}
\affiliation{Department of Physics, University of Maryland, College Park, MD, USA }
\author{J.P.~Harding}
\affiliation{Physics Division, Los Alamos National Laboratory, Los Alamos, NM, USA }
\author{S.~Hernandez}
\affiliation{Instituto de F\'{i}sica, Universidad Nacional Autónoma de México, Ciudad de Mexico, Mexico }
\author{B.~Hona}
\affiliation{Department of Physics and Astronomy, University of Utah, Salt Lake City, UT, USA }
\author{D.~Huang}
\affiliation{Department of Physics, Michigan Technological University, Houghton, MI, USA }
\author{F.~Hueyotl-Zahuantitla}
\affiliation{Universidad Autónoma de Chiapas, Tuxtla Gutiérrez, Chiapas, México}
\author{P.~Hüntemeyer}
\affiliation{Department of Physics, Michigan Technological University, Houghton, MI, USA }
\author{A.~Iriarte}
\affiliation{Instituto de Astronom\'{i}a, Universidad Nacional Autónoma de México, Ciudad de Mexico, Mexico }
\author{A.~Jardin-Blicq}
\affiliation{Max-Planck Institute for Nuclear Physics, 69117 Heidelberg, Germany}
\affiliation{Department of Physics, Faculty of Science, Chulalongkorn University, 254 Phayathai Road, Pathumwan, Bangkok 10330, Thailand}
\affiliation{National Astronomical Research Institute of Thailand (Public Organization), Don Kaeo, MaeRim, Chiang Mai 50180, Thailand}

\author{V.~Joshi}
\affiliation{Erlangen Centre for Astroparticle Physics, Friedrich-Alexander-Universit\"at Erlangen-N\"urnberg, Erlangen, Germany}
\author{D.~Kieda}
\affiliation{Department of Physics and Astronomy, University of Utah, Salt Lake City, UT, USA }
\author{A.~Lara}
\affiliation{Instituto de Geof\'{i}sica, Universidad Nacional Autónoma de México, Ciudad de Mexico, Mexico }
\author{W.H.~Lee}
\affiliation{Instituto de Astronom\'{i}a, Universidad Nacional Autónoma de México, Ciudad de Mexico, Mexico }
\author{J.~Lee}
\affiliation{University of Seoul, Seoul, Rep. of Korea}
\author{H.~León Vargas}
\affiliation{Instituto de F\'{i}sica, Universidad Nacional Autónoma de México, Ciudad de Mexico, Mexico }
\author{J.T.~Linnemann}
\affiliation{Department of Physics and Astronomy, Michigan State University, East Lansing, MI, USA }

\author{A.L.~Longinotti}
\affiliation{Instituto Nacional de Astrof\'{i}sica, Óptica y Electrónica, Puebla, Mexico }
\affiliation{Instituto de Astronom\'{i}a, Universidad Nacional Autónoma de México, Ciudad de Mexico, Mexico }
\author{G.~Luis-Raya}
\affiliation{Universidad Politecnica de Pachuca, Pachuca, Hgo, Mexico }
\author{J.~Lundeen}
\affiliation{Department of Physics and Astronomy, Michigan State University, East Lansing, MI, USA }
\author{K.~Malone}
\affiliation{Physics Division, Los Alamos National Laboratory, Los Alamos, NM, USA }
\author{V.~Marandon}
\affiliation{Max-Planck Institute for Nuclear Physics, 69117 Heidelberg, Germany}
\author{O.~Martinez}
\affiliation{Facultad de Ciencias F\'{i}sico Matemáticas, Benemérita Universidad Autónoma de Puebla, Puebla, Mexico }
\author{J.~Martínez-Castro}
\affiliation{Centro de Investigaci\'on en Computaci\'on, Instituto Polit\'ecnico Nacional, M\'exico City, M\'exico.}
\author{J.A.~Matthews}
\affiliation{Dept of Physics and Astronomy, University of New Mexico, Albuquerque, NM, USA }
\author{P.~Miranda-Romagnoli}
\affiliation{Universidad Autónoma del Estado de Hidalgo, Pachuca, Mexico }
\author{J.A.~Morales-Soto}
\affiliation{Universidad Michoacana de San Nicolás de Hidalgo, Morelia, Mexico }
\author{E.~Moreno}
\affiliation{Facultad de Ciencias F\'{i}sico Matemáticas, Benemérita Universidad Autónoma de Puebla, Puebla, Mexico }
\author{M.~Mostafá}
\affiliation{Department of Physics, Pennsylvania State University, University Park, PA, USA }
\author{A.~Nayerhoda}
\affiliation{Institute of Nuclear Physics Polish Academy of Sciences, PL-31342 IFJ-PAN, Krakow, Poland }
\author{L.~Nellen}
\affiliation{Instituto de Ciencias Nucleares, Universidad Nacional Autónoma de Mexico, Ciudad de Mexico, Mexico }
\author{M.~Newbold}
\affiliation{Department of Physics and Astronomy, University of Utah, Salt Lake City, UT, USA }
\author{M.U.~Nisa}
\affiliation{Department of Physics and Astronomy, Michigan State University, East Lansing, MI, USA }
\author{R.~Noriega-Papaqui}
\affiliation{Universidad Autónoma del Estado de Hidalgo, Pachuca, Mexico }
\author{L.~Olivera-Nieto}
\affiliation{Max-Planck Institute for Nuclear Physics, 69117 Heidelberg, Germany}
\author{N.~Omodei}
\affiliation{Department of Physics, Stanford University: Stanford, CA 94305–4060, USA}
\author{A.~Peisker}
\affiliation{Department of Physics and Astronomy, Michigan State University, East Lansing, MI, USA }
\author{Y.~Pérez Araujo}
\affiliation{Instituto de Astronom\'{i}a, Universidad Nacional Autónoma de México, Ciudad de Mexico, Mexico }
\author{E.G.~Pérez-Pérez}
\affiliation{Universidad Politecnica de Pachuca, Pachuca, Hgo, Mexico }
\author{C.D.~Rho}
\affiliation{University of Seoul, Seoul, Rep. of Korea}
\author{Y.J.~Roh}
\affiliation{University of Seoul, Seoul, Rep. of Korea}
\author{D.~Rosa-González}
\affiliation{Instituto Nacional de Astrof\'{i}sica, Óptica y Electrónica, Puebla, Mexico }
\author{E.~Ruiz-Velasco}
\affiliation{Max-Planck Institute for Nuclear Physics, 69117 Heidelberg, Germany}
\author{H.~Salazar}
\affiliation{Facultad de Ciencias F\'{i}sico Matemáticas, Benemérita Universidad Autónoma de Puebla, Puebla, Mexico }
\author{F.~Salesa Greus}
\affiliation{Institute of Nuclear Physics Polish Academy of Sciences, PL-31342 IFJ-PAN, Krakow, Poland }
\affiliation{Instituto de Física Corpuscular, CSIC, Universitat de València, E-46980, Paterna, Valencia, Spain}
\author{A.~Sandoval}
\affiliation{Instituto de F\'{i}sica, Universidad Nacional Autónoma de México, Ciudad de Mexico, Mexico }
\author{M.~Schneider}
\affiliation{Department of Physics, University of Maryland, College Park, MD, USA }
\author{H.~Schoorlemmer}
\affiliation{Max-Planck Institute for Nuclear Physics, 69117 Heidelberg, Germany}
\author{J.~Serna-Franco}
\affiliation{Instituto de F\'{i}sica, Universidad Nacional Autónoma de México, Ciudad de Mexico, Mexico }
\author{A.J.~Smith}
\affiliation{Department of Physics, University of Maryland, College Park, MD, USA }
\author{R.W.~Springer}
\affiliation{Department of Physics and Astronomy, University of Utah, Salt Lake City, UT, USA }
\author{P.~Surajbali}
\affiliation{Max-Planck Institute for Nuclear Physics, 69117 Heidelberg, Germany}
\author{M.~Tanner}
\affiliation{Department of Physics, Pennsylvania State University, University Park, PA, USA }
\author{K.~Tollefson}
\affiliation{Department of Physics and Astronomy, Michigan State University, East Lansing, MI, USA }
\author{I.~Torres}
\affiliation{Instituto Nacional de Astrof\'{i}sica, Óptica y Electrónica, Puebla, Mexico }
\author{R.~Torres-Escobedo}
\affiliation{Departamento de F\'{i}sica, Centro Universitario de Ciencias Exactase Ingenierias, Universidad de Guadalajara, Guadalajara, Mexico }
\affiliation{Tsung-Dao Lee Institute $\&$ School of Physics and Astronomy, Shanghai Jiao Tong University, Shanghai, China}
\author{R.~Turner}
\affiliation{Department of Physics, Michigan Technological University, Houghton, MI, USA }
\author{F.~Ureña-Mena}
\affiliation{Instituto Nacional de Astrof\'{i}sica, Óptica y Electrónica, Puebla, Mexico }
\author{L.~Villaseñor}
\affiliation{Facultad de Ciencias F\'{i}sico Matemáticas, Benemérita Universidad Autónoma de Puebla, Puebla, Mexico }
\author{T.~Weisgarber}
\affiliation{Department of Physics, University of Wisconsin-Madison, Madison, WI, USA }
\author{E.~Willox}
\affiliation{Department of Physics, University of Maryland, College Park, MD, USA }
\author{H.~Zhou}
\affiliation{Tsung-Dao Lee Institute $\&$ School of Physics and Astronomy, Shanghai Jiao Tong University, Shanghai, China}

\collaboration{HAWC Collaboration}
\correspondingauthor{Kelly Malone}
\email{kmalone@lanl.gov}
\begin{abstract}
The highest-energy known gamma-ray sources are all located within 0.5 degrees of extremely powerful pulsars. This raises the question of whether ultra-high-energy (UHE; $>$ 56 TeV) gamma-ray emission is a universal feature expected near pulsars with a high spin-down power. Using four years of data from the High Altitude Water Cherenkov (HAWC) Gamma-Ray Observatory, we present a joint-likelihood analysis of ten extremely powerful pulsars to search for subthreshold UHE gamma-ray emission correlated with these locations. We report a significant detection ($>$ 3$\sigma$), indicating that UHE gamma-ray emission is a generic feature of powerful pulsars. We discuss the emission mechanisms of the gamma rays and the implications of this result. The individual environment, such as the magnetic field and particle density in the surrounding area, appears to play a role in the amount of emission.
\end{abstract}

\section{Introduction}

Ultra-high-energy (UHE; $>$ 56 TeV) gamma-ray emission can be created via hadronic or leptonic processes. In the hadronic mechanism, a neutral pion decays into two gamma rays.  In the leptonic mechanism, a lower-energy photon scatters off a relativistic electron via inverse Compton scattering. The electron transfers part of its energy to the gamma ray, resulting in a higher-energy photon.

Traditionally, it was thought that UHE gamma-ray sources would be hadronic in nature, as leptonic emission is suppressed in this energy regime due to the Klein-Nishina (KN) effect.  However, present-day gamma-ray observatories have the sensitivity required to detect leptonic sources above 56 TeV. See, for example, the detections of the Crab Nebula as well as several known pulsar wind nebulae (PWN)~\citep{Crab,Amenomori2019,HECatalog,Abdalla2018}.  The astrophysical spectrum of a leptonic source typically exhibits significant curvature due to the KN effect~\citep{Moderski2005}. Conversely, hadronic sources follow the spectrum of their parent cosmic-ray population, which may or may not include a cutoff or curvature. 

The High Altitude Water Cherenkov (HAWC) Observatory is a gamma-ray observatory with an wide instantaneous field-of-view ($\sim$2 steradians) and sensitivity to energies between a few hundred GeV and a few hundred TeV. It is sensitive to sources with declinations between -26 and +64 degrees~\citep{Smith2015,Abeysekara2017a,Crab}.

The first HAWC catalog of UHE sources (\cite{HECatalog}; hereafter referred to as the ``eHWC" catalog) contains nine sources emitting above 56 TeV, three of which continue above 100 TeV.  The highest-energy sources all exhibit curvature in the spectrum.  Additionally, all nine sources are located within 0.5 degrees of pulsars. For eight of the nine sources, the pulsar has an extremely high spin-down power ($\dot{E} > 10^{36}$ erg/s). This is much higher than the number of high-$\dot{E}$ pulsars that would be expected to be found near UHE gamma-ray sources~\citep{HECatalog}.  Emission near pulsars is expected to be powered by a PWN or TeV halo and is therefore dominantly leptonic, even at the high energies studied here~\citep{2020arXiv201013960B,Sudoh2021}. While young pulsars are expected to have an associated supernova remnant (SNR) with accompanying hadronic emission, there has only been one detection of gamma-ray emission from an SNR to UHE, and leptonic emission from this source has not been conclusively ruled out~\citep{boomerang}. SNR theory starts to run into technical problems accelerating particles to these energies~\citep{2017AIPC.1792b0002G}. SNR acceleration to UHE energies may not be possible~\citep{snr2021}.

The proximity of these gamma-ray sources to the most powerful pulsars, along with the curvature in their spectra,  invites the question of whether UHE gamma-ray emission is a generic feature expected near these sources.  This is investigated in this paper through a joint-likelihood analysis of pulsars with $\dot{E} > 10^{36}$ erg/s to search for subthreshold sources in the HAWC data. While each source is too weak to be individually detected in HAWC's standard catalog search, analyzing the regions jointly may lead to a general detection for this source class. 

The paper is organized as follows: Section \ref{analysis} describes the details of the joint-likelihood method.  Section \ref{results} contains the results of the analysis. Section \ref{discussion} discusses implications of the results.

\section{Analysis method}\label{analysis}

\subsection{Joint-Likelihood Method}
In this paper, we perform a joint analysis of several high-$\dot{E}$ pulsars. The analysis uses a binned maximum-likelihood method that searches for a gamma-ray excess above the background, using a simple model that describes the UHE emission from pulsars based on various observables such as the distance and pulsar age. This is performed using the HAWC Accelerated Likelihood (HAL)\footnote{\url{https://github.com/threeML/hawc_hal}} plugin to the Multi-Mission Maximum Likelihood Framework (3ML)~\citep{Vianello2015}. We describe the background rejection and likelihood method in ~\cite{Crab}.  HAWC's background comes from two main sources: the Galactic diffuse background stemming from unresolved gamma-ray sources, and cosmic rays that are detected by HAWC and survive gamma/hadron separation cuts. Note that neither background is very large above 56 TeV. The Galactic diffuse background is largely due to Inverse Compton scattering, which is not very prominent at these energies. The fraction of cosmic rays surviving gamma/hadron separation cuts is also very low: $\sim$0.001~\citep{Abeysekara2017a}.

Data were collected over 1343 days between June 2015 and June 2019.  The data are binned in quarter-decade bins of reconstructed gamma-ray energy using the ``ground parameter" energy estimator, one of two energy estimators currently used by HAWC~\citep{Crab}. 

We use the last three quarter-decade energy bins from \cite{Crab}, restricting the analysis to reconstructed energies between 56 and 316 TeV. These are the highest energies probed by HAWC. At these energies, there is very little diffuse emission and less source confusion than at lower energies.

The spectral model is a power law:
\begin{equation}\label{eq:spectral}
\frac{dN}{dE} = A_i K \left( \frac{E}{E_0} \right)^{-\alpha},
\end{equation}
where $A_i$ accounts for the model-dependent relative flux of each source (see Section \ref{weights}), $K$ is the normalization, and $E_0$ is the pivot energy, which is fixed at the center of each energy bin. The three values used for $E_0$ in this analysis are 74.99 TeV, 133.35 TeV, and 237.14 TeV.  Setting the pivot energy at the center of each bin makes the analysis relatively insensitive to the choice of $\alpha$, which is fixed at 2.5.

We fit each source (see Table \ref{tab:pulsar}) individually in each energy bin, placing a step function at the boundaries of the bin to ensure that events that are mis-reconstructed in energy are excluded.  We assume the sources are spatially extended with a Gaussian morphology. The width is fixed at $\sigma$=0.34$^{\circ}$. The values of $\alpha$ and $\sigma$ are approximately the average values for the highest-energy gamma-ray sources from the eHWC catalog~\citep{HECatalog}. 

To obtain a joint-likelihood result, the log-likelihood profiles for each individual source, in each energy bin, are added and the value of $K$ (hereafter called $\hat{K}$) that optimizes this log-likelihood profile is found.  The total flux normalization, $\kappa$, for the ten sources combined is then just:

\begin{equation}\label{eq:k}
\kappa = \sum_{i=1}^{n_{sources}} \hat{K} A_i.
\end{equation}

Using this flux normalization, the total differential flux from all ten sources is:
\begin{equation}\label{eq:tot}
\frac{dN}{dE} = \kappa \left( \frac{E}{E_0} \right)^{-\alpha}.
\end{equation}

We calculate a test statistic (TS) to show how significant the value of $\kappa$ is: 
\begin{equation}
\mathrm{TS} = 2 \mathrm{ln} \frac{L_{S+B}(\kappa)}{L_B},
\end{equation}
where $L_{S+B}$ is the maximum likelihood for the signal-plus-background hypothesis and $L_{B}$ is the likelihood for the background-only hypothesis.  

After an overall best-fit value of $\kappa$ is determined, it is used as an input to a Markov chain Monte Carlo and a distribution of the value is determined. If the TS in a given energy bin is significant (TS  $> 4$), a Bayesian credible interval (68$\%$ containment) is shown, obtained from the estimated distribution of the parameter $\kappa$. The model only has one degree of freedom, so using Wilks Theorem~\citep{Wilks1938}, a TS value of 4 corresponds to 2$\sigma$.  Otherwise, 95$\%$ credible interval quasi-differential upper limit is plotted. A uniform prior is used in the Bayesian analysis. 

In the bins where an upper limit is determined, the range of expected upper limits are obtained from simulations of Poisson-fluctuated background. 

\subsection{Source Selection and Dataset}\label{sourceSel}

\begin{figure*}
\centering
\includegraphics[width=0.99\textwidth]{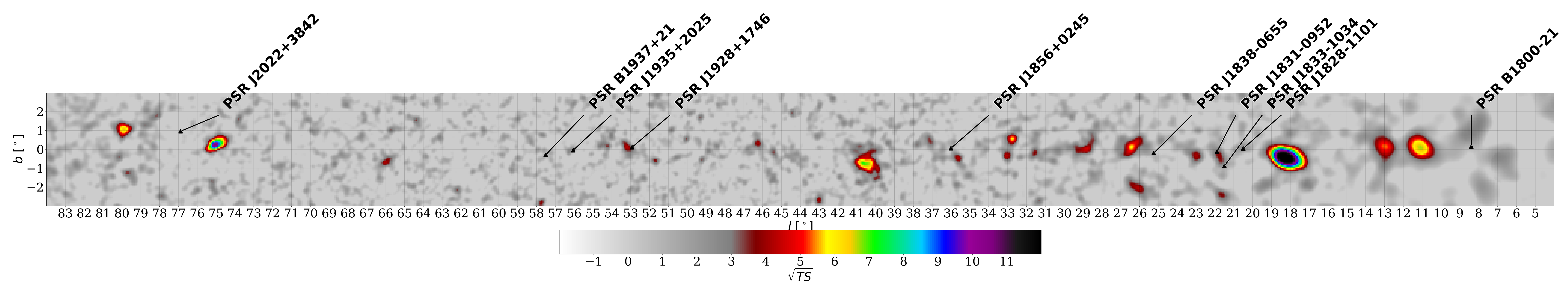}
\caption{$\sqrt{TS}$ for $\hat{E} >$ 56 TeV emission from HAWC, for the field-of-view used in this analysis. We assume a point source morphology. The locations of the ten sub-threshold pulsars used in this analysis are labeled.  }
\label{fig:gr56}
\end{figure*}

For this analysis, we define the source list by selecting all pulsars from the ATNF pulsar database, version 1.62\footnote{\url{https://www.atnf.csiro.au/research/pulsar/psrcat/}}~\citep{Manchester2005} in the inner Galactic plane that are within HAWC's field-of-view ($|b| < $ 1$^{\circ}$, 5$^{\circ}$ $< l <$ 90$^{\circ}$) and have $\dot{E} >$ 10${^{36}}$ erg/s. There are 24 pulsars that meet this criteria (see Table \ref{tab:pulsar}). Pulsars with a high $\dot{E}$ are centered around the Galactic plane, so the choice to concentrate on $|b| < 1^{\circ}$ only removes 3 additional pulsars from the analysis. We use this list of 24 pulsars to determine which models of emission are reasonable (see Section \ref{weights}).

After making this determination, this list is then downselected to search for sub-threshold gamma-ray sources. First, we remove all pulsars that are located within a degree of sources from the eHWC catalog~\citep{HECatalog}. This removes pulsars that already have significant UHE emission detected in their vicinity. Since the gamma-ray emission is modeled as extended in nature, this also removes pulsars whose associated emission may overlap with those known sources, which would require more detailed modeling.

We remove three additional pulsars from the source selection. PSR J1813-1749 is removed because HAWC has added more data since the publication of \cite{HECatalog} and has detected a new UHE source (eHWC-J1813-176) $\sim$0.2$^{\circ}$ away from the pulsar.   PSR J1913+1011 and PSR J1930+1852 have been removed because the known TeV emission in those regions is likely from a SNR, not a PWN~\citep{Abdalla2018a,Abdalla2018b}. Since the majority of the emission at these energies is expected to be from a PWN or TeV halo, this is done to prevent the introduction of a different, predominantly hadronic source class into the analysis. 

The final list is composed of ten pulsars that are candidates for sub-threshold gamma-ray emission. Figure \ref{fig:gr56} shows HAWC's $>$ 56 TeV map with these sources labeled.

\begin{table*}
\renewcommand{\arraystretch}{1.15}
\begin{center}
 \begin{tabular}{ | c || c| c | c | c| c | c| c | c| }
 \hline
 PSR name & RA ($^{\circ}$) & Dec ($^{\circ}$) & Age ($\frac{P}{2\dot{P}}$) (kyr) & $\dot{E}$ ($\times$10${^{36}}$ erg/s)  & Distance (kpc) & P (s) & $\dot{P}$ (ss$^{-1}$) & Subthreshold?   \\  
  \hline
B1800-21 &  270.96 & -21.62 & 15.8 & 2.2 & 4.40 & 0.134 & 1.35 $\times 10^{-13}$ & $\checkmark$ \\
J1809-1917 & 272.43 & -19.29 & 51.3 & 1.8 & 3.27 & 0.083 & 2.55 $\times 10^{-14}$ & \\
J1811-1925 & 272.87 & -19.42 & 23.3 & 6.4 & 5.00 & 0.065 & 4.4 $\times 10^{-14}$ & \\
J1813-1749 & 273.40 & -17.83 & 5.6 & 56 & 4.70 & 0.045 & 1.27 $\times 10^{-13}$ & \\
J1826-1256 & 276.54 & -12.94 & 14.4 & 3.6 & 1.55 & 0.110 & 1.21 $\times 10^{-13}$ & \\   
B1823-13 & 276.55 & -13.58 & 21.4 & 2.8 & 3.61 & 0.101 & 7.53 $\times 10^{-14}$ & \\
J1828-1101 & 277.08 & -11.03 & 77.1 & 1.6 & 4.77 & 0.072 & 1.48 $\times 10^{-14}$ & $\checkmark$ \\
J1831-0952 & 277.89 & -9.87 & 128 & 1.1 & 3.68 & 0.067 & 8.32 $\times 10^{-15}$ & $\checkmark$ \\
J1833-1034 & 278.39 & -10.57 & 4.85 & 34 & 4.10 & 0.062 & 2.02 $\times 10^{-13}$ & $\checkmark$ \\
J1837-0604 & 279.43 & -6.08 & 33.8 & 2.0 & 4.77 & 0.096 & 45.1 $\times 10^{-15}$ & \\
J1838-0537 & 279.73 & -5.62 & 4.89 & 6.0 & 2.0\footnote{Pseudo-distance from ~\cite{Pletsch2012}} & 0.146 & 4.72 $\times 10^{-13}$ & \\
J1838-0655 & 279.51 & -6.93 & 22.7 & 5.5 & 6.60 & 0.070 & 4.93 $\times 10^{-14}$ & $\checkmark$ \\
J1844-0346 & 281.14 & -3.78 & 11.6 & 4.2 & 2.40\footnote{Pseudo-distance derived from Eq. 3 of \cite{Wu2018}} & 0.113 & 1.55 $\times 10^{-13}$ &  \\
J1846-0258 & 281.60 & -2.98 & 0.728 & 8.1 & 5.8 & 0.327 & 7.11 $\times 10^{-12}$ & \\
J1849-0001 & 282.23 & -0.02 & 42.9 & 9.8 & 7.0\footnote{Distance estimate from \cite{Gotthelf2011}} & 0.039 & 1.42 $\times 10^{-14}$ & \\
J1856+0245 & 284.21 & 2.76 & 20.6 & 4.6 & 6.32 & 0.081 & 6.21 $\times 10^{-14}$ & $\checkmark$ \\
J1907+0602 & 286.98 & 6.04 & 19.5 & 2.8 & 2.37 & 0.107 & 8.68 $\times 10^{-14}$ & \\
J1913+1011 & 288.33 & 10.19 & 169 & 2.9 & 4.61 & 0.036 & 3.37 $\times 10^{-15}$ & \\
J1928+1746 & 292.18 & 17.77 & 82.6 & 1.6 & 4.34 & 0.069 & 1.32 $\times 10^{-14}$ & $\checkmark$ \\
J1930+1852 & 292.63 & 18.87 & 2.89 & 12 & 7.00 & 0.137 & 7.51 $\times 10^{-13}$ & \\
J1935+2025 & 293.92 & 20.43 & 20.9 & 4.7 & 4.60 & 0.080 & 6.08 $\times 10^{-14}$ & $\checkmark$ \\
B1937+21 & 294.91 & 21.58 & 2.35e5 & 1.1 & 3.50 & 0.002 & 1.05$\times 10^{-19}$ & $\checkmark$ \\
J2021+3651 & 305.27 & 36.85 & 17.2 & 3.4 & 1.8 & 0.104 & 9.57 $\times 10^{-14}$ & \\
J2022+3842 & 305.59 & 38.70 & 8.94 & 30 & 10.00 & 0.049 & 8.61 $\times 10^{-14}$ & $\checkmark$ \\
  \hline
\end{tabular}
\caption{Information on the pulsars. All information comes from the ATNF database, version 1.62~\citep{Manchester2005} except for some distance estimates which are not included in the pulsar database (see footnotes). Here, $\dot{E}$ is the spin-down energy loss rate, $P$ is the barycentric period, and $\dot{P}$ is the time derivative of the period. The checkmark in the last column denotes the ten pulsars included in the sub-threshold analysis.  }\label{tab:pulsar}
\end{center}
\end{table*}

\subsection{Models}\label{weights} 

The parameter $A_i$ in Equation \ref{eq:spectral} describes the relative contribution each pulsar receives in the analysis. This parameter can be used to test different models of gamma-ray emission near pulsars. For the models that rely on pulsar parameters, the relevant quantities are taken from the ATNF pulsar catalog, version 1.62~\citep{Manchester2005}. We consider a variety of different models.  In all descriptions, ``emission" refers solely to gamma-ray emission above 56 TeV. 
 
\begin{itemize}

\item \textbf{No model}: Here, $A_i$ for all sources is set equal to 1. All sources are treated equally and the emission is expected to be uncorrelated with pulsar parameters such as distance.

\item $\boldsymbol{1/d^2}$: In this model, $A_i$ is set to 1/$d^2$, where $d$ is the distance to the pulsar. This model assumes that closer sources will produce observable emission. 

\item $\boldsymbol{\dot{E}/d^2}$: Here, the 1/$d^2$ model discussed above is multiplied by the spin-down power of the pulsar. Therefore, closer, more-energetic pulsars have more gamma-ray emission.

\item \textbf{Inverse age}: In this model, $A_i$ is the inverse of the spin-down age. This is defined as $P/(2\dot{P})$, where $P$ and $\dot{P}$ are the period and the time derivative of the period, respectively. In this model, younger sources are more likely to have detectable emission. 

\item \textbf{Flux at 7 TeV}: Here, $A_i$ is the HAWC flux at 7 TeV. This model assumes that sources that are bright at multi-TeV energies should also give off detectable emission above 56 TeV. The 0.5$^\circ$ extended source map from the third catalog of HAWC sources (3HWC)~\citep{3hwc} is used to extract these values. $A_i$ is computed by averaging the flux from all pixels within a 0.5 degree radius of the pulsar.

\end{itemize}

\section{Results}\label{results}

We first run the joint-likelihood analysis (as described in Section \ref{analysis}) with the full list of 24 pulsars to determine which models of emission are reasonable.  Table \ref{tab:kValues} gives the total TS for each scenario. For each model, we can also calculate the expected gamma-ray flux above 56 TeV from an arbitrary pulsar using Equation \ref{eq:spectral}. The values of $K$, derived from Equation \ref{eq:k} are also given in Table \ref{tab:kValues}. In all cases, the TS is much higher than 5$\sigma$. Two scenarios, the flux at 7 TeV and 1/$d^2$, perform better than the ``no model case''. The model based on the inverse age of the pulsar performs significantly worse than the others, but the significance, $\sqrt{TS}$ is still well above the 5$\sigma$ level. 

\begin{table*}
\renewcommand{\arraystretch}{1.15}
\begin{center}
 \begin{tabular}{ | c || c| c | c | c | c| c| }
 \hline
Model &$K$ (56 $<$ E $<$ 100 TeV) & $K$ (100 $<$ E $<$ 177 TeV) & $K$ (177 $<$ E $<$ 316 TeV) & Units for $A_i$ & Total TS  \\
  \hline
No model & 2.55 $\times$ 10$^{-16}$ & 4.65 $\times$ 10$^{-17}$ & 9.46 $\times$ 10$^{-18}$ & dimensionless & 633      \\
1/$d^2$ & 2.44 $\times$ 10$^{-16}$ & 4.50 $\times$ 10$^{-16}$ & 8.12 $\times$ 10$^{-17}$ & kpc$^{-2}$ & 734    \\
$\dot{E}/d^2$ & 4.06 $\times$ 10$^{-13}$ & 6.88 $\times$ 10$^{-14}$ & 1.04 $\times$ 10$^{-14}$ & erg m$^{-2}$ s$^{-1}$ & 568  \\
Inverse age & 6.36 $\times$ 10$^{-13}$ & 1.10 $\times$ 10$^{-13}$ & 1.65 $\times$ 10$^{-14}$ & yr$^{-1}$ & 152  \\
Flux at 7 TeV & 5.42 $\times$ 10$^{-3}$ & 9.11 $\times$ 10$^{-4}$ & 1.90 $\times$ 10$^{-4}$ & TeV$^{-1}$ cm$^{-2}$ s$^{-1}$ & 886   \\
  \hline
\end{tabular}
\caption{The proportionality constants needed to calculate the expected gamma-ray flux near a given pulsar. The dimensions for $A_iK$ are energy$^{-1}$ distance$^{-1}$ time$^{-1}$.  With the value of $K$ known and $A_i$ in the units given by the last column, the expected gamma-ray flux can easily be calculated using Equation \ref{eq:spectral}. $K$ is reported at the pivot energy in each bin. Proportionality constants are desrived from the full sample of 24 pulsars. }  \label{tab:kValues}
\end{center}
\end{table*}

Satisfied that the chosen models are adequate, we then run the joint-likelihood analysis using the downselected list of ten pulsars (as described in Section \ref{sourceSel}) to search for sub-threshold leptonic emission from pulsars. Results are given in Table \ref{tab:TS}. Each of the models give a total TS value above 9, corresponding to a detection of more than 3$\sigma$. For some models, the TS is much higher. The 1/$d^2$ model gives the highest TS: 41.3 (6.4$\sigma$).  The same two models as before, 1/$d^2$ and the gamma-ray flux at 7 TeV, perform better than the ``no model'' case, but in this case 1/$d^2$ performs the best, with a TS slightly higher than the ``gamma-ray flux at 7 TeV'' scenario.

\begin{table*}
\renewcommand{\arraystretch}{1.15}
\begin{center}
 \begin{tabular}{ | c || c| c | c | c |  }
 \hline
 Model & TS (56 $<$ E $<$ 100 TeV) & TS (100 $<$ E $<$ 177 TeV) & TS (177 $<$ E $<$ 316 TeV) & Total TS \\
  \hline
No model & 27.9 & 8.33 & 1.59 & 37.9   \\
1/$d^2$ & 31.9 & 9.08 & 1.29 & 41.3 \\
$\dot{E}$/$d^2$ & 9.58 & 5.24 & 0.00 & 14.8 \\
Inverse age & 9.19 & 3.79 & 0.03 & 13.0\\
Flux at 7 TeV & 26.3 & 9.61 & 3.62 & 39.6 \\
  \hline
\end{tabular}
\caption{The test statistic for the joint-likelihood analysis for each model, using the ten sub-threshold sources. Note that $\kappa$ is fit individually in each energy bin so the last column is not a joint TS. }\label{tab:TS}
\end{center}
\end{table*}

\begin{figure}
\centering
\includegraphics[width=0.48\textwidth]{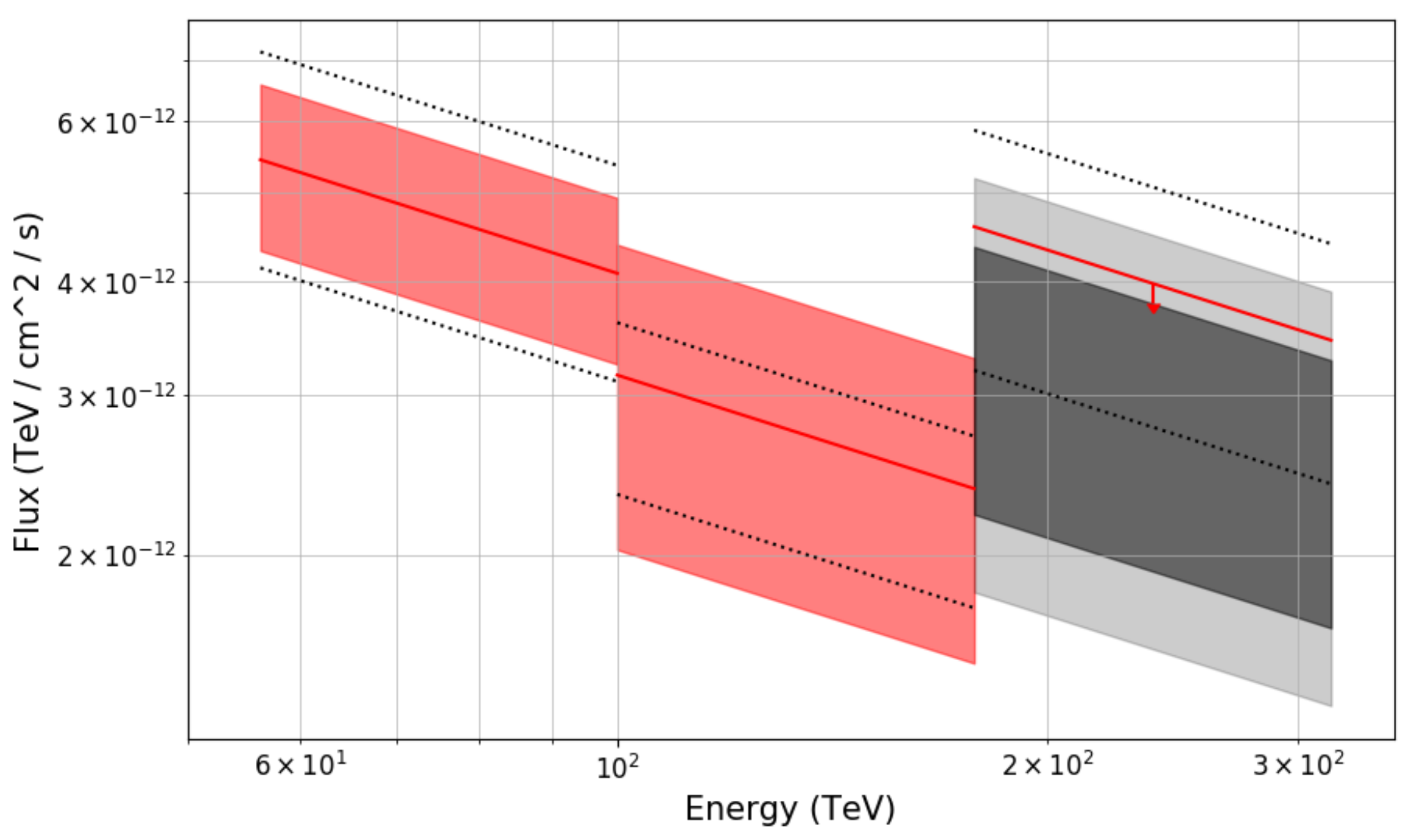}
\caption{The total combined flux from the ten sub-threshold candidates for the scenario where the pulsars have a relative contribution defined by 1/$d^2$, where $d$ is the distance between the pulsar and the Earth. For the first two energy bins, the TS $>$ 4 so Bayesian credible intervals (68$\%$ containment) are plotted. In the last energy bin, there is no significant detection so a 95$\%$ upper limit is plotted. The dark and light grey bands are 68$\%$ and 90$\%$ containment for expected upper limits from Poisson fluctuated background. The dotted lines are systematic uncertainties on the central value (i.e., the solid red line may be as low or as high as the dotted line once systematic uncertainties are included, see section \ref{sec:systematics}) }
\label{fig:fluxWeight}
\end{figure}

Figure \ref{fig:fluxWeight} shows the total flux from all ten sub-threshold candidates for this best-case scenario. The figures for the other models can be seen in the Appendix.  Table \ref{tab:flux}, also located in the Appendix, gives the total flux normalizations for each model.

The Appendix also includes a discussion of how often TS this high would be expected from stacking random points on the sky. 

\subsection{Systematic Uncertainties}\label{sec:systematics}

Systematic uncertainties are broken down into two categories: detector systematics and modeling systematics. Detector systematics, described in \cite{Crab}, stem from mis-modeling of detector quantities such as the photomultiplier tube  threshold and charge in simulated Monte Carlo events.  Each systematic is treated independently; the results are added in quadrature to get a total uncertainty. This process is repeated in each energy bin. Depending on the energy bin and model assumed, the detector systematic ranges from 10$\%$ to 25$\%$.  

Modeling systematic uncertainties investigate how the analysis would change if the emission near the sub-threshold  pulsars is different from what is assumed in the main analysis. Several modeling systematics are considered:
\begin{itemize}
\item Spectral indices in the power law ($\alpha$ in Equation \ref{eq:spectral}) of 2.0 and 3.0 
\item Replacing the power-law spectral model (Equation \ref{eq:spectral}) with a power-law with an exponential cutoff: 
\begin{equation}\label{eq:spectral2}
\frac{dN}{dE} = A_i K \left( \frac{E}{E_0} \right)^{-\alpha} e^{-E/E_{cut}}.
\end{equation}
Here, $\alpha$ is fixed at 2.5 and $E_{cut}$ is fixed at 60 TeV, which are average values from~\cite{HECatalog}
\item Changing the Gaussian width to 0.23$^{\circ}$ and 0.45$^{\circ}$. These are $\pm$ 1 standard deviation from the average extension of the sources from HAWC's eHWC catalog~\citep{HECatalog}.
\end{itemize}

The modeling systematics are larger than the detector systematics, driven predominantly by the assumed source size.  In the last energy bin, assuming a power law with an exponential cutoff instead of a power law is also a dominant effect. 

Depending on energy and model, the modeling systematic ranges from 13$\%$ to 34$\%$. The sum of the detector and modeling systematics, added in quadrature, are denoted with a dotted black line in Figure \ref{fig:fluxWeight} and in Figures \ref{fig:unweighted} through \ref{fig:age} in the Appendix.

As described in ~\cite{3hwc}, the absolute pointing uncertainty of HAWC is declination-dependent and may be larger than previously thought; perhaps as large as 0.3$^{\circ}$ at the edges of the field-of-view. This means that the TS values presented in this paper may be underestimated; the size of this effect is $\sim$10$\%$. The flux may also be underestimated. The flux underestimate is a function of energy and ranges from 9$\%$ in the 56-100 TeV bin, decreasing to a negligible amount (0.6$\%$) in the 177-316 TeV bin.

\section{Discussion}\label{discussion}

Regardless of which model is used, it is clear that the areas around high-$\dot{E}$ pulsars show hints of emitting at ultra-high energies ($>$ 56 TeV).  Interestingly, some models based on the pulsar parameters, such as $\dot{E}$ or inverse age, have much lower TS values than the ``no model" case, implying that they do not describe the emission well.  Two models that perform better than the ``no model" case are the gamma-ray flux at 7 TeV and 1/$d^2$.

It is unclear at this time why some pulsar parameters do not seem to be good predictors of emission. Some of these parameters have fairly large uncertainties, which could be a contributing factor. For example, the characteristic age of the pulsar can differ from the true age.

However, the uncertainties on the pulsar distances are relatively small. For the ATNF pulsar database, 95$\%$ of pulsars will have a relative error of less than a factor of 0.9 in their distance estimate (10$\%$ uncertainty)~\citep{pulsarDist}.

Alternatively, the individual environment that each pulsar is in could play a large factor in the amount of  UHE emission. While the pulsar itself is the particle accelerator, diffusion of electrons and positrons away from the pulsar is strongly dependent on quantities such as the density and magnetic field of the environment.  Note that this may be only true for the high-$\dot{E}$ pulsars studied here, which are relatively young; this means the magnetic fields are likely to be affected by the pulsar age and SNR interactions. For weaker, older pulsars, the magnetic field density and environment are instead likely dominated by the ISM. 

Also, the emission mechanisms are still uncertain. While it is commonly assumed that the bulk of emission from the vicinity of a pulsar is from a PWN and therefore predominantly leptonic, a hadronic contribution cannot be \textit{a priori} discarded. While emission from associated SNRs are unlikely, some have raised the possibility of more exotic hadronic emission mechanisms in or near PWN~\citep{hadron1,hadron2}. Several of the HAWC sources known to emit above 56 TeV, most notably eHWC J1908+063 and eHWC J1825-134, have molecular clouds nearby~\citep{Duvidovich2020,Voisin2016}.  These molecular clouds may be serving as a target for a portion of the gamma-ray production.

Multi-wavelength and multi-messenger campaigns can help disentangle emission mechanisms.  A neutrino detection coincident with one of these pulsars would be a smoking gun for hadronic emission mechanisms in or near PWN. However, a recent stacked analysis looking for neutrino emission from PWN by IceCube did not yield a detection~\citep{IceCube2020}. 

Electromagnetic multi-wavelength studies could also be helpful. A leptonic source emitting above 56 TeV will have a different signature at lower energies than a hadronic one.  For example, 100 TeV gamma rays imply a synchrotron peak in the keV regime, assuming a 3 $\mu$G field~\citep{Hinton2009}, with the emission extending up to the MeV energy range.  If the emission is instead predominantly hadronic, there will be no such peak at these energies.  Proposed experiments such as AMEGO~\citep{McEnery} will be important in distinguishing emission mechanisms. 

\section{Conclusions} 

In this study, we have searched for UHE gamma-ray emission in the vicinity of pulsars with an $\dot{E} > 10^{36}$ erg/s.  We find, with high significance ($>$ 3$\sigma$) , that UHE gamma-ray emission is a generic feature in the vicinity of this class of pulsars.  1/$d^2$ is the model that gives the highest TS; a source that is closer to us is more likely to have observed UHE emission. Other pulsar parameters do not seem to be good predictors of emission. This implies that the environments the pulsars are located in may play a role in the amount of emission. 

The TS values obtained are higher than would be expected from combining random points in the Galactic plane. There is the possibility that all known gamma-ray sources emit above this energy threshold, albeit at an extremely low flux.
 
Multimessenger and multiwavelength studies are needed to disentagle the origin of the UHE emission from these pulsars. 

\acknowledgments

We acknowledge the support from: the US National Science Foundation (NSF); the US Department of Energy Office of High-Energy Physics; the Laboratory Directed Research and Development (LDRD) program of Los Alamos National Laboratory; Consejo Nacional de Ciencia y Tecnolog\'ia (CONACyT), M\'exico, grants 271051, 232656, 260378, 179588, 254964, 258865, 243290, 132197, A1-S-46288, A1-S-22784, c\'atedras 873, 1563, 341, 323, Red HAWC, M\'exico; DGAPA-UNAM grants IG101320, IN111315, IN111716-3, IN111419, IA102019, IN110621; VIEP-BUAP; PIFI 2012, 2013, PROFOCIE 2014, 2015; the University of Wisconsin Alumni Research Foundation; the Institute of Geophysics, Planetary Physics, and Signatures at Los Alamos National Laboratory; Polish Science Centre grant, DEC-2017/27/B/ST9/02272; Coordinaci\'on de la Investigaci\'on Cient\'ifica de la Universidad Michoacana; Royal Society - Newton Advanced Fellowship 180385; Generalitat Valenciana, grant CIDEGENT/2018/034; Chulalongkorn University’s CUniverse (CUAASC) grant. Thanks to Scott Delay, Luciano D\'iaz and Eduardo Murrieta for technical support.

\bibliography{stackedBibliography}

\appendix

\subsection{Additional joint-likelihood results}
\begin{figure}[h]
\centering
\includegraphics[width=0.48\textwidth]{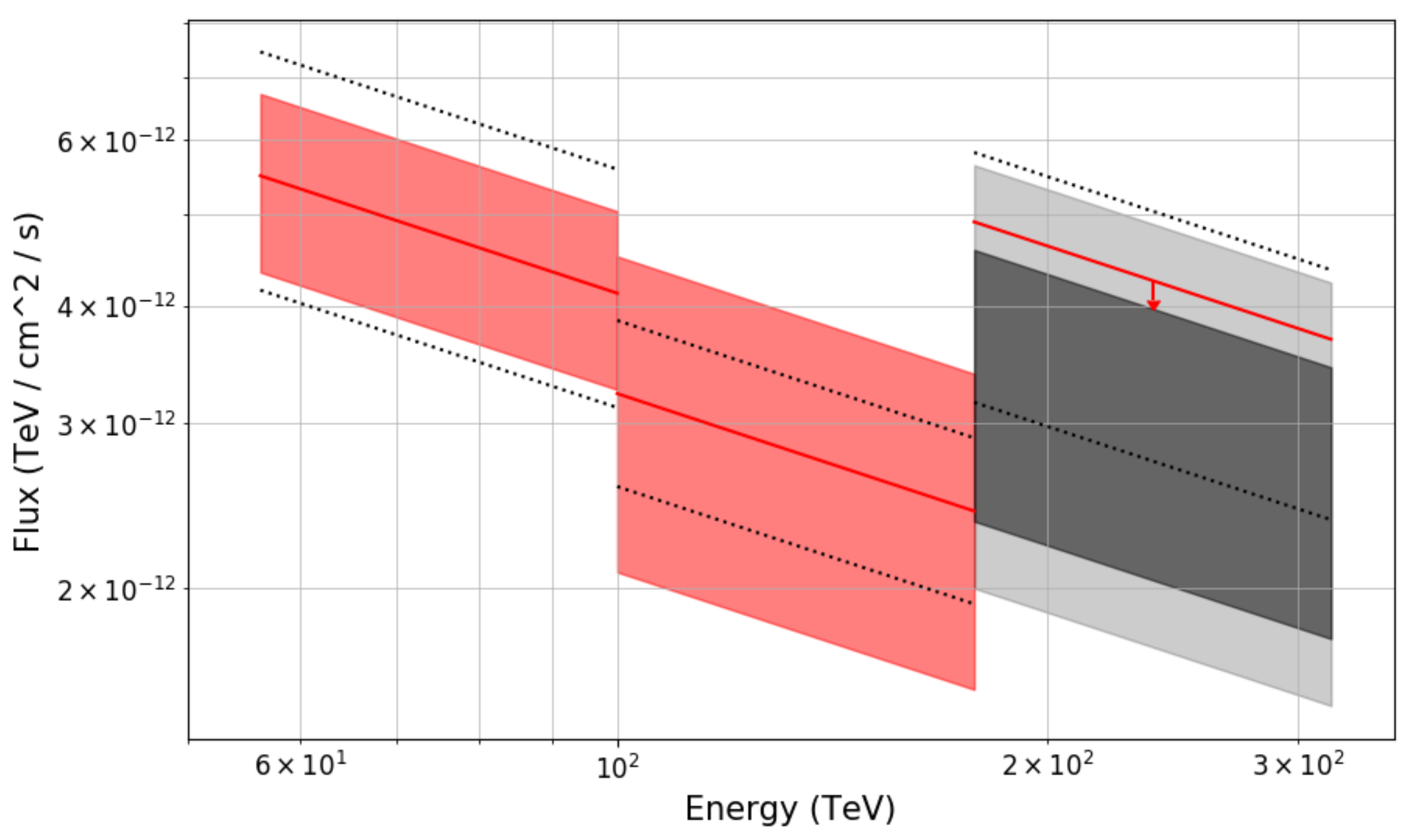}
\caption{Identical to Figure \ref{fig:fluxWeight} from the main text, but for the ``no model" case. }
\label{fig:unweighted}
\end{figure}

\begin{figure}
\centering
\includegraphics[width=0.48\textwidth]{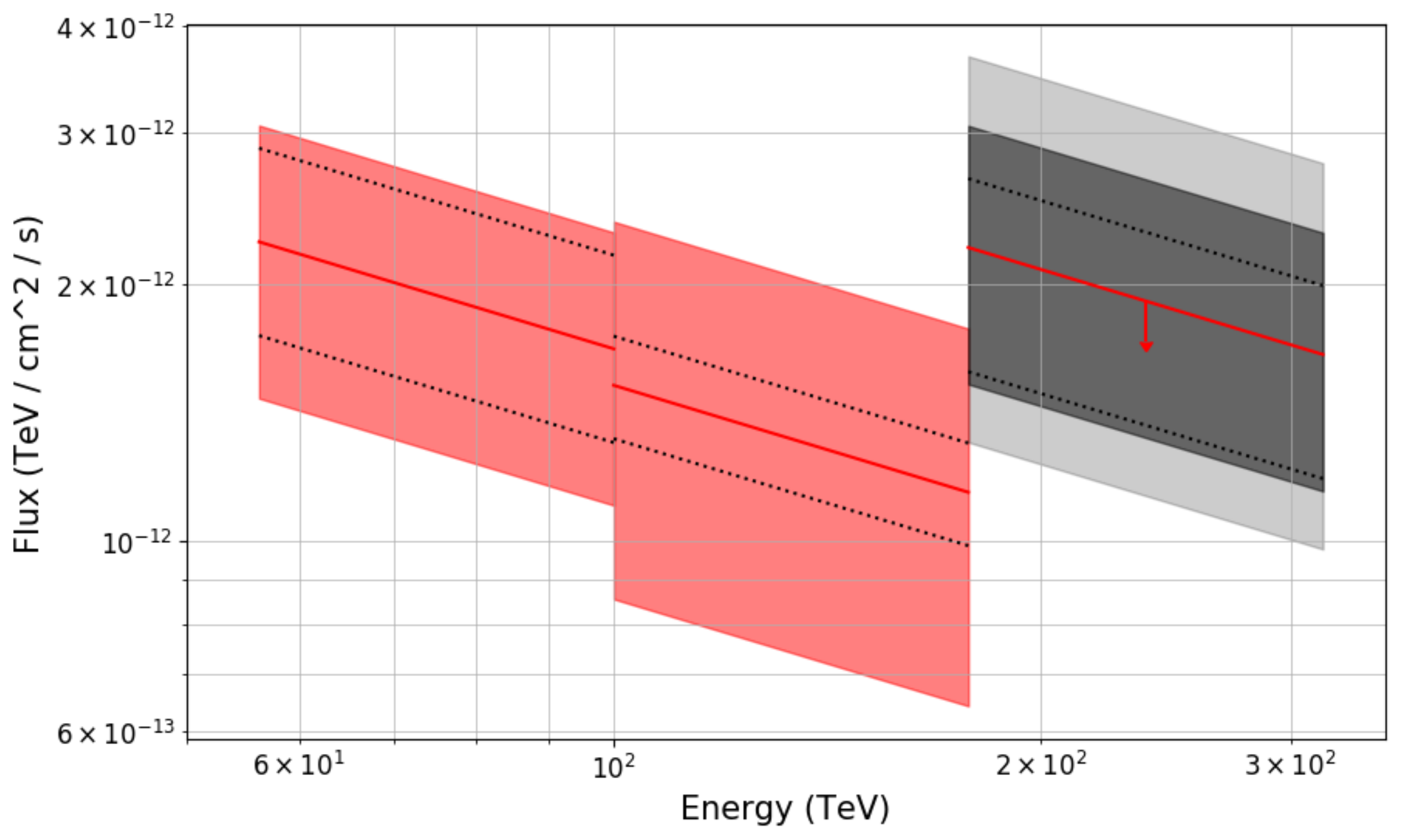}
\caption{Identical to Figure \ref{fig:fluxWeight} from the main text, but for the $\dot{E}/d^2$ model.  }
\label{fig:edot}
\end{figure}

\begin{figure}
\centering
\includegraphics[width=0.48\textwidth]{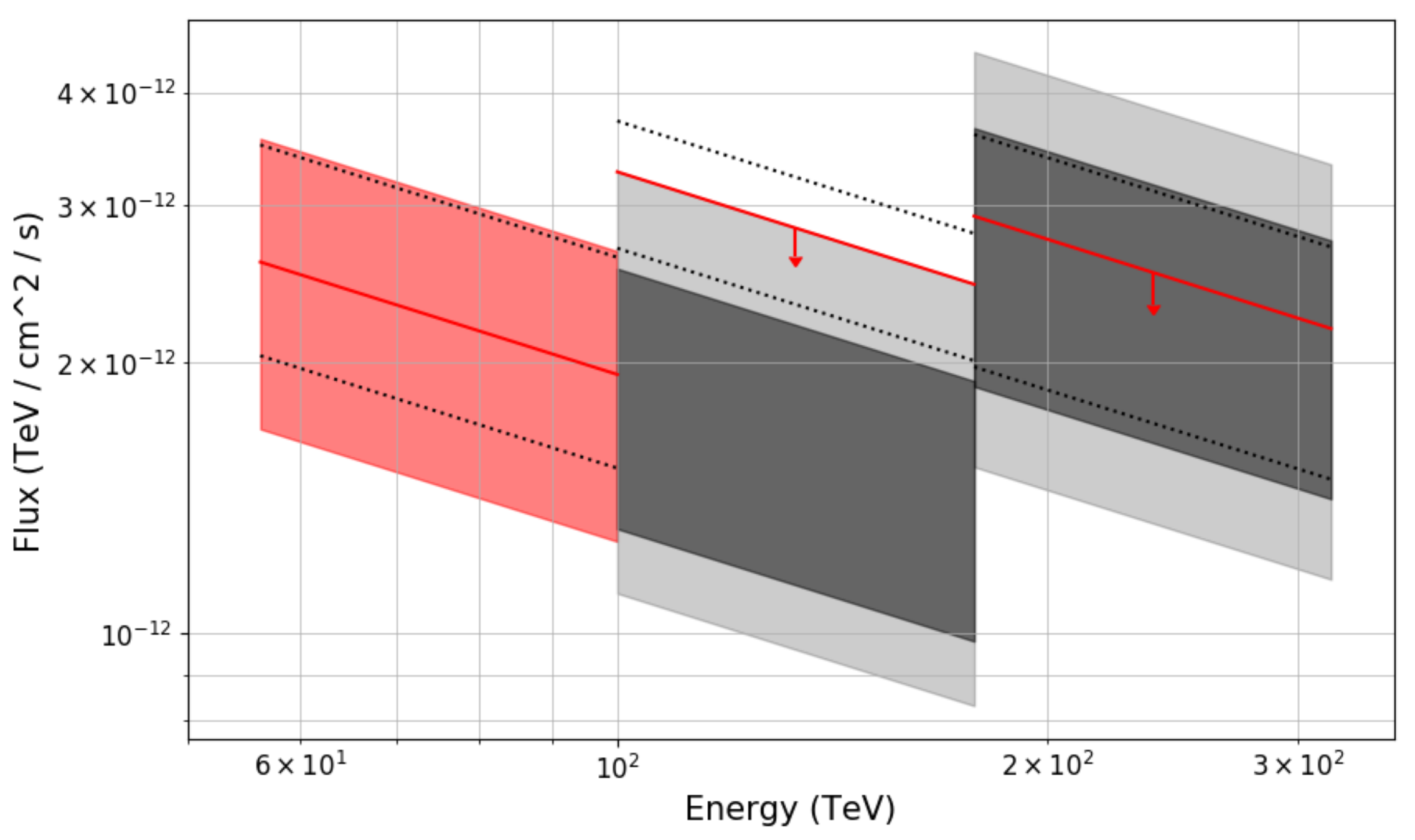}
\caption{Identical to Figure \ref{fig:fluxWeight} from the main text, but for the inverse age model. }
\label{fig:age}
\end{figure}

Figures \ref{fig:unweighted} through \ref{fig:age} are analogous to Figure \ref{fig:fluxWeight} in the main text, but for the other four models that have been investigated (described in Section \ref{weights}). All figures contain the combined flux for the joint-likelihood analysis of the ten sub-threshold pulsars.   Table \ref{tab:flux} contains the total combined flux normalizations for each of the models that have been considered. 

\begin{figure}
\centering
\includegraphics[width=0.48\textwidth]{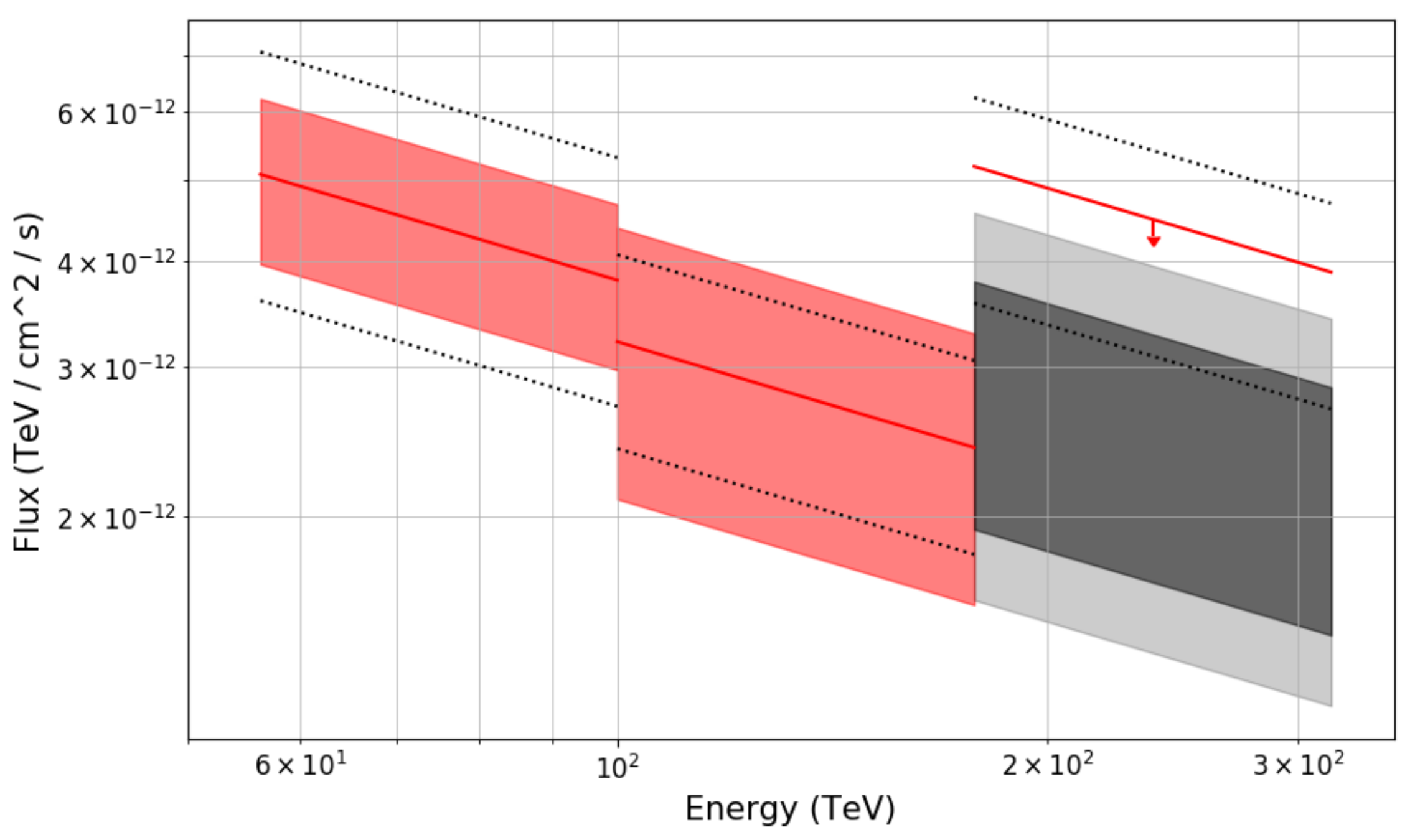}
\caption{Identical to Figure \ref{fig:fluxWeight} from the main text, but for the model defined by the gamma-ray flux at 7 TeV. }
\label{fig:d2}
\end{figure}

\begin{table*}
\renewcommand{\arraystretch}{1.15}
\begin{center}
 \begin{tabular}{ | c || c| c | c | c |  }
 \hline
 Model & $\kappa$ (56 $<$ E $<$ 100 TeV) & $\kappa$ (100 $<$ E $<$ 177 TeV) & $\kappa$ (177 $<$ E $<$ 316 TeV)  \\
  \hline
No model & 8.47${_{-1.78}^{+1.88}}$ $\times$ 10$^{-16}$ & 1.57$_{-0.56}^{+0.63}$ $\times$ 10$^{-16}$ & 7.56 $\times$ 10$^{-17}$    \\
1/$d^2$ & 8.38$_{-1.72}^{+1.76}$ $\times$ 10$^{-16}$ & 1.54$_{-0.55}^{+0.60}$ $\times$ 10$^{-16}$ & 7.07 $\times$ 10$^{-17}$   \\
$\dot{E}$/$d^2$ & 3.45$_{-1.18}^{+1.27}$ $\times$ 10$^{-16}$ & 7.41$_{-3.24}^{+4.10}$ $\times$ 10$^{-17}$ & 3.39 $\times$ 10$^{-17}$ \\
Inverse age & 3.99$_{-1.39}^{+1.48}$ $\times$ 10$^{-16}$ & 1.59 $\times$ 10$^{-16}$ & 4.48 $\times$ 10$^{-17}$ \\
Flux at 7 TeV & 7.80$_{-1.69}^{+1.78}$ $\times$ 10$^{-16}$ & 1.56$_{-0.55}^{+0.57}$ $\times$ 10$^{-16}$ & 7.98 $\times$ 10$^{-17}$  \\
  \hline
\end{tabular}
\caption{The total combined flux normalization for the sub-threshold joint-likelihood analysis for each of the models ($\kappa$ from Equation \ref{eq:tot}). The units are TeV$^{-1}$ cm$^{-2}$ s$^{-1}$. The flux normalization is reported at the pivot energy, which is the center of each bin. Values without uncertainties are upper limits, otherwise the values correspond to the 68$\%$ containment for the Bayesian credible interval. Uncertainties are statistical only.}  \label{tab:flux}
\end{center}
\end{table*}

\subsection{Testing of Random Backgrounds}

We investigate how often randomly chosen non-source locations give TS values as high as in the study presented in the main text. Sets of ten randomly chosen points from the analysis region (\mbox{$4^{\circ}$ $< l <$ 90$^{\circ}$; $|b| < 1^{\circ}$}) are run through the joint-likelihood analysis.  Points that are within a degree of either a known high-energy source or any of the selected ATNF pulsars are excluded. Because the gamma-ray emission is assumed to be spatially extended, this is necessary to avoid contributions from the known sources and the pulsars of interest.

This analysis is performed for 2,877 sets of ten random source locations. Only five of these randomly chosen sets have a TS greater than 37.9. This is the TS from the sub-threshold joint-likelihood ``no model" case and is used as the comparison because the other models require known pulsar information. This means that the results that we have obtained are significant at the 3$\sigma$ level. 

We also investigate sets of ten randomly chosen sources from HAWC's third catalog (3HWC)~\citep{3hwc} to see if any of HAWC's previously detected TeV sources emit at UHE. Once again, sources that are known to emit above 56 TeV and sources within a degree of the ATNF pulsars used in the nominal analysis are removed.  Note that the majority of HAWC sources are leptonic in origin~\citep{2017PhRvD..96j3016L}.

Due to the relatively small number of remaining 3HWC sources after this downselection (48 of the original 65 3HWC sources), it is not possible to run thousands of sets of ten sources without repeating a large subset of the sources. We instead run 100 sets of ten randomly chosen 3HWC sources. 

None of these 100 trials have a TS above the value from the sub-threshold joint-likelihood unmodeled case. The highest TS from this study is 26.9 and the mean is 12.2 (standard deviation 5.5). This implies that known gamma-ray sources that are not located near high-$\dot{E}$ pulsars are unlikely to emit at high energies, or if they do, their fluxes are very low and combining ten sources is not enough for a detection.

We also explore combining 20 3HWC sources at a time, instead of the ten that were combined in the preceding paragraph. The entire TS distribution is shifted to higher values. The highest TS is 45.9 (higher than in the ``no model" case); the mean is 24.6 and the standard deviation is 7.2. The higher TS values when more 3HWC sources are combined implies that UHE emission may be a generic feature of known gamma-ray sources, but with a very low flux. This should be investigated, but may be hard to do with current-generation experiments. Proposed experiments such as SWGO~\citep{SWGO}, will be important here.

\end{document}